\documentclass[two column, prA]{revtex4}%
\usepackage{amsmath}
\usepackage{graphicx}
\usepackage{amsfonts}
\usepackage{amssymb}
\usepackage{color}%
\providecommand{\U}[1]{\protect\rule{.1in}{.1in}}
\ifx\pdfoutput\relax\let\pdfoutput=\undefined\fi
\newcount\msipdfoutput
\ifx\pdfoutput\undefined\else
\ifcase\pdfoutput\else
\ifx\paperwidth\undefined\else
\ifdim\paperheight=0pt\relax\else\pdfpageheight\paperheight\fi
\ifdim\paperwidth=0pt\relax\else\pdfpagewidth\paperwidth\fi
\fi\fi\fi
\begin{document}
\title[Quantum oscillator model of photons]{The quantum oscillator model of electromagnetic excitations revisited\\ }
\author{Margaret Hawton}
\affiliation{Department of Physics, Lakehead University, Thunder Bay, ON, Canada, P7B 5E1}
\email{mhawton@lakeheadu.ca}

\begin{abstract}
We revisit the quantum oscillator model of the electromagnetic field and
conclude that, while the nonlocal positive and negative frequency ladder
operators generate a photon Fock basis, the Hermitian field operators obtained
by second quantization of real Maxwell fields describe photon-antiphoton pairs
that couple locally to Fermionic matter and can be modeled classically. Their
commutation relations define a scalar product that can be the basis of a first
quantized theory of single photons. Since a one-photon state collapses to a
zero-photon state when the photon is counted, the field describing it must be
interpreted as a probability amplitude.

\end{abstract}
\maketitle

\section{Introduction}

Quantum field theory (QFT) is the basis for our understanding of energy and
matter in flat space-time. In QFT classical fields are second quantized and
particles are defined as discrete excitations of these fields. In the
electromagnetic (EM) case these discrete field excitations are called photons.
Photons have integral helicity $\lambda=\pm1$ which makes them bosons whose
second quantized fields satisfy commutation relations. Since there is no
exclusion principle for bosons, an EM state is an arbitrary linear combination
of $n$-photon states where $n$ is any positive integer or $0$, that is it is a
whole number. The mathematical description of this ladder of photon states is
borrowed from the theory of massive quantum oscillators in which a negative
frequency operator creates a excitation and positive frequency operator
annihilates one.

According to the Hegerfeldt theorem \cite{Hegerfeldt} a positive (or negative)
frequency function localized in a finite region spreads instantaneously
throughout space and according to the Reeh-Schleider theorem of algebraic QFT
\cite{RS} there are no local annihilation or creation operators. Defining the
positive anti-local frequency operator as $\Omega\equiv c\left(  -\nabla
^{2}\right)  ^{1/2}$ where $-\nabla^{2}$ is the Laplacian, the second order EM
wave equation $\left(  \partial_{t}^{2}+\Omega^{2}\right)  \mathbf{A}_{\perp
}=0$ can be factored into $\left(  i\partial_{t}+\Omega\right)  \left(
-i\partial_{t}+\Omega\right)  \mathbf{A}_{\perp}^{\pm}=0$ to give first order
equations for nonlocal positive and negative frequency fields. By all these
criteria, single photon states are nonlocal.

Under the operation of charge conjugation, C, every particle is carried into
its antiparticle. Since photons are their own antiparticles Charge-Parity-Time
(CPT) symmetry requires that they be represented by real fields and Hermitian
field operators \cite{GellMann}. The Fermion four-current is odd under charge
conjugation since electrons are exchanged with positrons. To maintain
invariance of the current-field interaction and the Dirac equation the photon
four-potential should also be odd under charge conjugation. If $A_{j}^{+}$ is
a positive frequency photon four-potential and $A_{j}^{-}=A_{j}^{+\ast}$
describes a negative frequency antiphoton, QFT requires that $A_{j}=\left(
A_{j}^{+}-A_{j}^{-}\right)  /\sqrt{2}i$. Only this odd combination of photon
and antiphoton fields couples to Fermionic matter. 

In this manuscript we implement the QFT oscillator model of photon states in
detail and discuss its consequences.\ Our calculation was motivated by the
second quantized theory of Federico and Jauslin \cite{FJ} that in turn was
inspired by the work of Bialynicki-Birula \cite{BB}. Experiment and QFT tell
us that only whole numbers of photons exist, there are no fractional photons.
In the next section we spell out the mathematical details of the EM oscillator
model with an emphasis on its relationship to Maxwell's equations (MEs). In
the Conclusion we propose a physical model consistent with QFT, experiment and
the oscillator model described in Section II and discuss first quantization,
the probability interpretation and the classical EM field.

\section{Second quantization}

The analogy with a massive oscillator is sufficiently precise that we include
a summary of its Schr{\"{o}}dinger quantum mechanics. For an oscillator with
mass $m$, position $x$ with time derivative $\overset{\cdot}{x}$ and momentum
$p=m\overset{\cdot}{x}$ the Lagrangian is $L=\frac{1}{2}m\overset{\cdot
}{x}^{2}-\frac{1}{2}m\omega^{2}x^{2}+fx$ where $f$ is the external driving
force. The momentum conjugate to $x$ is $p=\partial\mathcal{L}/\partial
\overset{\cdot}{x}=m\overset{\cdot}{x}$, the equation of motion is
$\overset{\cdot}{p}=-m\omega^{2}x+f$ and the Hamiltonian is $H=\frac{p^{2}%
}{2m}+\frac{1}{2}m\omega^{2}x^{2}-fx$. For a free classical oscillator subject
to the initial condition $x\left(  0\right)  =0$, $x\left(  t\right)
=\omega^{-1}$ $\overset{\cdot}{x}$ $\left(  0\right)  \sin\omega t$ and
$p\left(  t\right)  =m\overset{\cdot}{x}\left(  0\right)  \cos\omega t$ so
$x+\frac{i}{m\omega}p=i\omega^{-1}\overset{\cdot}{x}\left(  0\right)
\exp\left(  -i\omega t\right)  $ is a positive frequency function. To
quantize, position and momentum are raised to the status of operators
satisfying the commutation relations $\left[  \widehat{x},\widehat{p}\right]
=i\hbar$ where $h=2\pi\hbar$ is Planck's constant. The oscillator states
$\left\vert n\right\rangle $ then have energies $E_{n}=\hbar\omega\left(
n+\frac{1}{2}\right)  $ where $n$ is a whole number. The nonhermitian
operators $\widehat{a}=\sqrt{\frac{m\omega}{2\hbar}}$ $\left(  \widehat{x}%
+\frac{i}{m\omega}\widehat{p}\right)  $ and $\widehat{a}^{\dag}=\sqrt
{\frac{m\omega}{2\hbar}}\left(  \widehat{x}-\frac{i}{m\omega}\widehat{p}%
\right)  $ that satisfy the commutation relation $\left[  \widehat{a}%
,\widehat{a}^{\dag}\right]  =1$ annihilate and create energy $\hbar\omega$ on
the ladder of oscillator states. These commutation relations can be used to
prove that $\widehat{a}^{\dag}\left\vert n\right\rangle =\sqrt{n+1}\left\vert
n+1\right\rangle $ and $\widehat{a}\left\vert n\right\rangle =\sqrt
{n}\left\vert n-1\right\rangle $. The normalized $n$-phonon eigenstates of the
homogeneous oscillator equation with energies $E_{n}$ are $\left\vert
n\right\rangle =\frac{\widehat{a}^{\dag n}}{\sqrt{n!}}\left\vert
0\right\rangle $ where $\left\vert 0\right\rangle $ is the oscillator ground
state. The ladder operators define a Fock phonon basis but do not provide a
mechanism for transition between physical states - that requires an external
driving force.

Since QFT is based on classical fields, a classical EM Lagrangian density,
$\mathcal{L}$, is sufficient. To avoid the need for nonlocal separation of the
longitudinal and transverse fields and ensure gauge independence it will be
assumed that the Fermionic matter is globally charge neutral \cite{CT}. Two
Lagrangians differing by a total time derivative describe the same system.
Since $\varepsilon_{0}\mathbf{E}_{\parallel}=-\mathbf{P}_{\parallel}$ in a
globally charge neutral system\textbf{,}
\begin{equation}
\mathbf{D\equiv}\varepsilon_{0}\mathbf{E}+\mathbf{P}_{\parallel}%
=\varepsilon_{0}\mathbf{E}_{\perp}\label{D}%
\end{equation}
is transverse. When written in terms of the electric and magnetic fields%
\begin{align}
\mathbf{E} &  =-\partial_{t}\mathbf{A}-\nabla\phi,\;\label{E}\\
\mathbf{B} &  =\nabla\times\mathbf{A,}\label{B}%
\end{align}
the Standard EM Lagrangian density is equivalent to
\begin{align}
\mathcal{L} &  =\frac{1}{2}\varepsilon_{0}\left(  \mathbf{E^{2}-}%
c^{2}\mathbf{B}^{2}\right)  +\mathbf{j}\cdot\mathbf{A}-\frac{d}{dt}\left(
\mathbf{P}_{\parallel}\cdot\mathbf{A}\right)  \nonumber\\
&  =\frac{1}{2}\left(  \mathbf{D\cdot E}_{\perp}\mathbf{-}\varepsilon_{0}%
c^{2}\mathbf{B}^{2}\right)  +\mathbf{j}_{\perp}\cdot\mathbf{A}_{\perp
}.\label{L}%
\end{align}
$\text{The Lagrangian is then }L=\int d\mathbf{x}\mathcal{L}\left(
t,\mathbf{x}\right)  $. Here $\mathbf{A}_{\perp}\left(  t,\mathbf{x}\right)  $
plays the role of a position coordinate in the oscillator analogy and the
momentum conjugate to $\mathbf{A}_{\perp}$ is $\Pi=-\mathbf{D}$. The
permittivity $\varepsilon_{0}$ takes the place of mass in the oscillator
analogy and the external driving force is $\mathbf{j}_{\perp}$. Two of the
classical MEs have been used to define the four-potential $A$. The remaining
two classical MEs are%
\begin{align}
\nabla\cdot\mathbf{D}\  &  =0,\quad\nonumber\\
\partial_{t}\mathbf{D} &  =\varepsilon_{0}c^{2}\bigtriangledown\times
\mathbf{B}-\mathbf{j}_{\perp}\label{MEs}%
\end{align}
where the transverse current density is%
\begin{equation}
\mathbf{j}_{\perp}=\partial_{t}\mathbf{P}_{\perp}+\mathbf{\nabla\times
M,}\label{jperp}%
\end{equation}
$\mathbf{M}$ being the magnetization. In Eqs. (\ref{D}) to (\ref{jperp}) the
transverse current is treated as an external driving force. Substitution of
(\ref{B}) into (\ref{MEs}) gives%
\begin{equation}
\partial_{t}\mathbf{D}-\varepsilon_{0}\Omega^{2}\mathbf{A}_{\perp}%
=-\mathbf{j}_{\perp}\label{ME}%
\end{equation}
where $\Omega^{2}$ is a local operator and%
\begin{equation}
\Omega=c\left(  -\nabla^{2}\right)  ^{1/2}\label{omega}%
\end{equation}
is the nonlocal, self-adjoint, positive, invertible frequency operator
\cite{GettingAcquainted}. The transverse gauge independent spring-like
restoring force is $\varepsilon_{0}c^{2}\bigtriangledown\times
\mathbf{B=\varepsilon}_{0}\Omega^{2}\mathbf{A}_{\perp}$.

The oscillator-like waves described by (\ref{ME}) propagate and lead to a Fock
basis of states containing a whole number of photons. Following the definition
of the ladder operators for a massive oscillator with $\widehat{x}%
\rightarrow\widehat{\mathbf{A}}_{\perp}$ and $\widehat{p}\rightarrow
-\widehat{\mathbf{D}}$ the $xp$-commutator becomes%
\begin{equation}
\left[  \widehat{\mathbf{A}}_{\perp},\cdot\widehat{\mathbf{D}}\right]
=-i\hbar\label{ADcommutation}%
\end{equation}
where\ the commutator of three-vectors is defined as
\begin{equation}
\left[  \widehat{\mathbf{V}}_{1},\cdot\widehat{\mathbf{V}}_{2}\right]
\equiv\widehat{\mathbf{V}}_{1}\cdot\widehat{\mathbf{V}}_{2}%
-\widehat{\mathbf{V}}_{2}\cdot\widehat{\mathbf{V}}_{1}.\label{Vcomm}%
\end{equation}
The second quantized positive frequency annihilation operator analogous to
$\widehat{a}$ is%
\begin{equation}
\widehat{\boldsymbol{\psi}}=\sqrt{\frac{\varepsilon_{0}}{2\hbar}}\Omega
^{1/2}\left[  \widehat{\mathbf{A}}_{\perp}-i\left(  \varepsilon_{0}%
\Omega\right)  ^{-1}\widehat{\mathbf{D}}\right]  .\label{psi_op}%
\end{equation}
Powers of the operators, $\Omega^{s}$, do not commute with the field
operators; instead their commutator with a field operator is $\left[
\Omega^{s},\widehat{\mathbf{V}}\right]  =\left(  \Omega^{s}\widehat{\mathbf{V}%
}\right)  $ in which $\Omega^{s}$ acts only on the field $\widehat{\mathbf{V}%
}$. It is nontrivial but straightforward to prove using $\left[
\widehat{A}\widehat{B},\widehat{C}\widehat{D}\right]  =\left[  \widehat{A}%
,\widehat{C}\right]  \widehat{B}\widehat{D}+\widehat{C}\left[  \widehat{A}%
,\widehat{D}\right]  \widehat{B}+\widehat{A}\widehat{C}\left[  \widehat{B}%
,\widehat{D}\right]  +\widehat{A}\left[  \widehat{B},\widehat{C}\right]
\widehat{D}$ that (\ref{psi_op}) satisfies the commutation relation%
\begin{equation}
\left[  \widehat{\boldsymbol{\psi}},\cdot\widehat{\boldsymbol{\psi}}^{\dagger
}\right]  =1.\label{comm}%
\end{equation}

Expression (\ref{psi_op}) satisfies the Schr\"{o}dinger equation (SE)
\begin{equation}
i\partial_{t}\widehat{\boldsymbol{\psi}}=\Omega\widehat{\boldsymbol{\psi}%
}-\left(  2\varepsilon_{0}\hslash\Omega\right)  ^{-1/2}\widehat{\mathbf{j}%
}_{\perp}.\label{SE}%
\end{equation}
Its real and imaginary parts,%
\begin{align}
\widehat{\mathbf{D}} &  =-\varepsilon_{0}\partial_{t}\widehat{\mathbf{A}%
}_{\perp},\label{EAt}\\
\partial_{t}\widehat{\mathbf{D}}-\varepsilon_{0}\Omega^{2}\widehat{\mathbf{A}%
}_{\perp} &  =-\widehat{\mathbf{j}}_{\perp}\label{MEop}%
\end{align}
are a second quantized version of Maxwell's equations. Substitution of
(\ref{EAt}) in (\ref{psi_op}) gives $\widehat{\boldsymbol{\psi}}=\sqrt
{\frac{2\varepsilon_{0}}{\hbar}}\widehat{\mathbf{A}}_{\perp}^{+}$, verifying
that it is indeed a positive frequency operator. The current density operator
$\widehat{\mathbf{j}}_{\perp}$ can be classical or an operator if atomic
transitions are taken into account. Substitution of (\ref{EAt}) in
(\ref{MEop}) gives%
\begin{equation}
\partial_{ct}^{2}\widehat{\mathbf{A}}_{\perp}-\nabla^{2}\widehat{\mathbf{A}%
}_{\perp}=-\mu_{0}\widehat{\mathbf{j}}_{\perp}.\label{2ndorder}%
\end{equation}

The zero-photon state, $\left\vert 0\right\rangle $, satisfies
$\widehat{\boldsymbol{\psi}}\left\vert 0\right\rangle =0$ and the one-photon
states are%
\begin{equation}
\left\vert \boldsymbol{\psi}\right\rangle =\widehat{\boldsymbol{\psi}%
}^{\dagger}\left\vert 0\right\rangle .\label{OnePhoton}%
\end{equation}
The commutation relation%
\begin{equation}
\left[  \widehat{\boldsymbol{\psi}},\cdot\widehat{\boldsymbol{\psi}}%
^{\prime\dagger}\right]  =\left\langle \boldsymbol{\psi}|\boldsymbol{\psi
}^{\prime}\right\rangle \label{compsi}%
\end{equation}
generalizes (\ref{comm}). Its zero-photon expectation value is
\begin{equation}
\left\langle 0\right\vert \left[  \widehat{\boldsymbol{\psi}},\cdot
\widehat{\boldsymbol{\psi}}^{\prime\dagger}\right]  \left\vert 0\right\rangle
=\int d\mathbf{x}\boldsymbol{\psi}^{\ast}\left(  t,\mathbf{x}\right)
\cdot\boldsymbol{\psi}^{\prime}\left(  t,\mathbf{x}\right)
\label{ScalarProduct}%
\end{equation}
where $d\mathbf{x}$ is an infinitesimal volume in three-dimensional space. By
inspection of (\ref{psi_op}) and (\ref{ScalarProduct}) the one-photon wave
function is%
\begin{equation}
\boldsymbol{\psi}\left(  t,\mathbf{x}\right)  =\sqrt{\frac{\varepsilon_{0}%
}{2\hbar}}\Omega^{1/2}\left[  \mathbf{A}_{\perp}\left(  t,\mathbf{x}\right)
-i\left(  \varepsilon_{0}\Omega\right)  ^{-1}\mathbf{D}\left(  t,\mathbf{x}%
\right)  \right]  \label{Psi}%
\end{equation}
where $\left\langle \psi|\psi\right\rangle =1$ for a physical one-photon
state. The photon number density is%
\begin{equation}
\boldsymbol{\psi}^{\ast}\cdot\boldsymbol{\psi}=\frac{1}{2\hbar}\left[
\left\vert \left(  \varepsilon_{0}\Omega\right)  ^{1/2}\mathbf{A}_{\perp
}\right\vert ^{2}+\left\vert \left(  \varepsilon_{0}\Omega\right)
^{-1/2}\mathbf{D}\right\vert ^{2}\right]  .\label{ndensity}%
\end{equation}
This corresponds to $H/\hbar\omega$ in the oscillator model. There is a
nonlocal relationship between photon number density and energy \cite{FJ}. The
zero-photon expectation value of the position-momentum commutation relation is%
\begin{equation}
\frac{i}{\hbar\varepsilon_{0}}\left\langle 0\right\vert \left[
\widehat{\mathbf{A}},\cdot\widehat{\mathbf{D}}^{\prime\dagger}\right]
\left\vert 0\right\rangle =\frac{1}{2}\left(  \left\langle \boldsymbol{\psi
}|\boldsymbol{\psi}^{\prime}\right\rangle +\left\langle \boldsymbol{\psi
}^{\prime}|\boldsymbol{\psi}\right\rangle \right)  .\label{A,.D}%
\end{equation}

Plane waves form a basis that will allow us to perform calculations in
$\mathbf{k}$-space. Here we will follow the derivation of Fock space in
\cite{SZ} by starting with the periodic boundary conditions $k_{i}L=2\pi
n_{i}$ for $i=x,y,z$ in volume $V=L^{3}$ and then taking the $V\rightarrow
\infty$ limit. Covariant normalization will be used to give fields of the
classical form. The annihilation operator for vector $\mathbf{k}$ with
helicity $\lambda$ will be written as%
\begin{align}
\widehat{\boldsymbol{\psi}}_{\lambda\mathbf{k}}\left(  t,\mathbf{x}\right)
&  =\mathbf{e}_{\lambda}\left(  \mathbf{k}\right)  \widehat{a}_{\lambda
\mathbf{k}}\text{,}\label{psi_op_k}\\
\widehat{a}_{\lambda\mathbf{k}} &  =\frac{1}{V^{1/2}}\omega_{k}^{1/2}%
e^{-i\left(  \omega_{k}t-\mathbf{k}\cdot\mathbf{x}\right)  }\label{a_op_k}%
\end{align}
for mutually orthogonal transverse polarization unit vectors $\mathbf{e}%
_{\lambda}\left(  \mathbf{k}\right)  =\left(  \mathbf{e}_{\theta}%
+i\lambda\mathbf{e}_{\phi}\right)  /\sqrt{2}$, $\lambda=\pm1$,$\ $where
$\mathbf{e}_{\theta},$ $\mathbf{e}_{\phi}$ and $\mathbf{e}_{\mathbf{k}}$ are
orthonormal $\mathbf{k}$-space spherical polar unit vectors. These discrete
plane waves satisfy $\left\langle \boldsymbol{\psi}_{\lambda\mathbf{k}%
}|\boldsymbol{\psi}_{\lambda^{\prime}\mathbf{k}^{\prime}}\right\rangle
=\delta_{\lambda\lambda^{\prime}}\delta_{\mathbf{kk}^{\prime}}$ and the
commutation relations
\begin{equation}
\left[  \widehat{a}_{\lambda\mathbf{k}},\widehat{a}_{\lambda^{\prime
}\mathbf{k}^{\prime}}^{\dag}\right]  =\delta_{\lambda\lambda^{\prime}}%
\delta_{\mathbf{k,k}^{\prime}}.\label{k_commutation}%
\end{equation}
It can be verified using the commutation relations (\ref{k_commutation}) that
\begin{equation}
\left\vert a_{\lambda\mathbf{k}n}\right\rangle \equiv\frac{\left[
\widehat{a}_{\lambda\mathbf{k}}\right]  ^{n}}{\sqrt{n!}}\left\vert
0\right\rangle \label{psi_n}%
\end{equation}
are the normalized $n$-photon Fock states. The number of states per unit
volume is lim$_{V\rightarrow\infty}\Delta\mathbf{n}/V=d\mathbf{k}/\left(
2\pi\right)  ^{3}$ where $\Delta\mathbf{n}$ is the number of states so
$V^{-1}\sum_{\mathbf{k}}\rightarrow\left(  2\pi\right)  ^{-3}\int%
_{t}d\mathbf{k}$ and the invariant scalar product%

\begin{align}
\left\langle \boldsymbol{\psi}_{\lambda\mathbf{k}}|\boldsymbol{\psi}%
_{\lambda^{\prime}\mathbf{k}^{\prime}}\right\rangle  & =\delta_{\lambda
\lambda/}\int d\mathbf{xe}^{i\left(  \mathbf{k-k}^{\prime}\right)
\cdot\mathbf{x}}\nonumber\\
& =\delta_{\lambda\lambda/}\left(  2\pi\right)  ^{3}\omega_{k}\delta\left(
\mathbf{k-k}^{\prime}\right)  \label{kcomm}%
\end{align}
defines a basis of orthonormal states. In this basis the covariant positive
frequency vector potential operator describing a general EM mode is%
\begin{equation}
\widehat{\mathbf{A}}_{\lambda}^{+}=\sqrt{\frac{\hbar}{\varepsilon_{0}}}%
\int\frac{d\mathbf{k}}{\left(  2\pi\right)  ^{3}\omega_{k}}c_{\lambda}\left(
\mathbf{k}\right)  \widehat{a}_{\lambda}\left(  \mathbf{k}\right)
\mathbf{e}_{\lambda}\left(  \mathbf{k}\right)  \label{A+}%
\end{equation}
\ where $c_{\lambda}\left(  \mathbf{k}\right)  $ is the invariant probability
amplitude for $\left(  \mathbf{k},\lambda\right)  $,%
\begin{align}
\widehat{\mathbf{A}}_{\lambda}^{-}  & =\widehat{\mathbf{A}}_{\lambda}^{+\dag
},\ \widehat{\mathbf{A}}_{\perp}=\sum_{\lambda=\pm1}\left(
\widehat{\mathbf{A}}_{\lambda}^{+}+\widehat{\mathbf{A}}_{\lambda}^{-}\right)
,\label{A+-}\\
\widehat{\mathbf{D}}  & =-\varepsilon_{0}\partial_{t}\widehat{\mathbf{A}%
}_{\perp},\ \widehat{\mathbf{B}}=\mathbf{\nabla\times}\widehat{\mathbf{A}%
}_{\perp},\label{DandB}%
\end{align}
and $\widehat{\boldsymbol{\psi}}$ is given by (\ref{psi_op}). The plane wave
basis has the advantage that states can be described by the scalar
$c_{\lambda}\left(  \mathbf{k}\right)  $. The photon annihilation operator is%
\begin{equation}
\widehat{\boldsymbol{\psi}}=\sum_{\lambda=\pm1}\int\frac{d\mathbf{k}}{\left(
2\pi\right)  ^{3}\omega_{k}^{1/2}}c_{\lambda}\left(  \mathbf{k}\right)
\widehat{a}_{\lambda}\left(  \mathbf{k}\right)  \mathbf{e}_{\lambda}\left(
\mathbf{k}\right)  .\label{psiFourier}%
\end{equation}
Eqs. (\ref{kcomm}) and (\ref{psiFourier}) then give the $\mathbf{k}$-space
form of the scalar product as%
\begin{equation}
\left\langle \widehat{\boldsymbol{\psi}}|\widehat{\boldsymbol{\psi}}%
^{\prime\dagger}\right\rangle =\sum_{\lambda=\pm1}\int\frac{d\mathbf{k}%
}{\left(  2\pi\right)  ^{3}\omega_{k}}c_{\lambda}^{\ast}\left(  \mathbf{k}%
\right)  c_{\lambda}^{\prime}\left(  \mathbf{k}\right)  .\label{commFourier}%
\end{equation}
The one-photon fields can be normalized using (\ref{ScalarProduct}) or
(\ref{commFourier}) and used as the basis for a first quantized theory of
single photons. Consistent with the nonlocal relationship between fields and
number amplitude, expressions for number density and scalar product and
one-boson fields in \cite{Mostafazadeh} were based on the nonlocal sign of
frequency operator $\widehat{\epsilon}=\Omega^{-1}\partial_{ct}$.

A general $n$-photon state is%

\begin{equation}
\left\vert \widehat{\boldsymbol{A}}_{n}\right\rangle =\sqrt{\frac{\hbar
}{\varepsilon_{0}}}\sum_{\lambda=\pm1}\int\frac{d\mathbf{k}}{\left(
2\pi\right)  ^{3}\omega_{k}}c_{\lambda}\left(  \mathbf{k}\right)
\mathbf{e}_{\lambda}\left(  \mathbf{k}\right)  \left\vert a_{\lambda
\mathbf{k}n}\right\rangle .\label{nphotonstate}%
\end{equation}
The response to a classical current density $\mathbf{j}_{\perp}\left(
t,\mathbf{x}\right)  $ is the coherent state%
\begin{equation}
\left\vert a_{\lambda\boldsymbol{k}\alpha}\right\rangle =e^{-\left\vert
\alpha_{\lambda\mathbf{k}}\right\vert ^{2}/2}\sum_{n=0}^{\infty}\frac{\left(
\alpha_{\lambda\mathbf{k}}\right)  ^{n}}{\sqrt{n!}}\left\vert a_{\lambda
\mathbf{k}n}\right\rangle \label{Coherent}%
\end{equation}
where the probability to count a photon with helicity $\lambda$ and wavevector
$\mathbf{k}$ is $\alpha_{\lambda\mathbf{k}}^{\ast}\alpha_{\lambda\mathbf{k}}$
and \cite{SZ}%
\begin{equation}
\alpha_{\lambda\mathbf{k}}=\sqrt{\frac{\hbar}{\varepsilon_{0}}}\int_{0}%
^{t}\int dt^{\prime}\int d\mathbf{x}c_{\lambda}\left(  \mathbf{k}\right)
\mathbf{e}_{\lambda}\left(  \mathbf{k}\right)  \cdot\mathbf{j}_{\perp}\left(
t,\mathbf{x}\right)  e^{-ikx}.\label{alpha}%
\end{equation}
\begin{equation}
k\equiv\left(  \omega_{k}/c,\mathbf{k}\right)  ,\ x\equiv\left(
ct,\mathbf{x}\right)  \text{ and }kx\equiv\omega_{k}t-\mathbf{k}%
\cdot\mathbf{x}\label{kx}%
\end{equation}
is a Lorentz invariant. For a system in a coherent state defined by the set
\begin{equation}
\left\vert \left\{  \alpha_{\lambda\mathbf{k}}\right\}  \right\rangle
=\sum_{\lambda}%
{\displaystyle\prod_{\mathbf{k}}}
\left\vert \alpha_{\lambda\mathbf{k}}\right\rangle \label{CoherentState}%
\end{equation}
the expectation value of (\ref{A+-}),%
\begin{equation}
\mathbf{A}_{\lambda}\left(  t,\mathbf{x}\right)  =\sqrt{\frac{\hbar
}{\varepsilon_{0}}}\int\frac{d\mathbf{k}}{\left(  2\pi\right)  ^{3}\omega_{k}%
}\alpha_{\lambda\mathbf{k}}\mathbf{e}_{\lambda}\left(  \mathbf{k}\right)
e^{-ikx}+cc,\label{Alambda}%
\end{equation}
is real.

A two level system in a linear combination of ground state $\left\vert
g\right\rangle $ and excited state $\left\vert e\right\rangle $ provides a
non-zero current equal to the expectation value of the electron current
operator, $\left\langle -e\widehat{p}/m\right\rangle .$ If the EM state is a
linear combination of a zero photon state, $\left\vert 0\right\rangle $, and a
one-photon state, $\left\vert 1\right\rangle $, the atom and field form a pair
of coupled oscillators.

The choice%
\begin{equation}
c_{\lambda}\left(  \mathbf{k}\right)  \equiv c_{x\lambda}\left(
\mathbf{k}\right)  =e^{-ikx}\label{Localized}%
\end{equation}
substituted in (\ref{ScalarProduct}) gives
\begin{equation}
\left\langle 0\left\vert \left[  \widehat{\boldsymbol{\psi}}_{\lambda x}%
,\cdot\widehat{\boldsymbol{\psi}}_{\lambda x^{\prime}}^{\dagger}\right]
\right\vert 0\right\rangle =\left\langle \boldsymbol{\psi}_{\lambda
x}|\boldsymbol{\psi}_{\lambda x^{\prime}}\right\rangle
\label{PhotonPropagation}%
\end{equation}
that describes propagation from space-time point $x^{\prime}=\left(
t^{\prime},\mathbf{x}^{\prime}\right)  $ to $x=\left(  t,\mathbf{x}\right)  $
while (\ref{A,.D}) gives%
\begin{align}
\frac{1}{2\varepsilon_{0}\hbar}\left\langle 0\left\vert \left[
\widehat{\mathbf{A}}_{\lambda x},\cdot\widehat{\mathbf{D}}_{\lambda x^{\prime
}}\right]  \right\vert 0\right\rangle  &  =\frac{1}{2}\int\frac{d\mathbf{k}%
}{\left(  2\pi\right)  ^{3}}\left[  e^{ik\left(  x-x^{\prime}\right)
}\right.  \nonumber\\
&  \left.  -e^{ik\left(  x-x^{\prime}\right)  }\right]
,\label{PhotonAntiphoton}%
\end{align}
describes propagation of a photon from $x^{\prime}$ to $x$ plus propagation of
a photon backwards in time from $x$ to $x^{\prime}$ which is equivalent to
propagation of a photon from $x^{\prime}$ to $x$ plus propagation of an
antiphoton from $x^{\prime}$ to $x$. While (\ref{PhotonPropagation}) is
complex for $t\neq t$ and spreads instantaneously, (\ref{PhotonAntiphoton}) is
real and propagates causally \cite{Validation}. On the $t=t^{\prime}$
hyperplane%
\begin{equation}
\frac{1}{2\varepsilon_{0}\hbar}\left\langle 0\left\vert \left[
\widehat{\mathbf{A}}_{\lambda\left(  t,\mathbf{x}\right)  },\cdot
\widehat{\mathbf{D}}_{\lambda\left(  t,\mathbf{x}^{\prime}\right)  }\right]
\right\vert 0\right\rangle =\delta_{\lambda\lambda^{\prime}}\delta\left(
\mathbf{x-x}^{\prime}\right)  \label{xBasis}%
\end{equation}
provides an orthonormal basis of localized states. The operators
$\widehat{\mathbf{A}}_{\lambda\left(  t,\mathbf{x}\right)  }$ and
$\widehat{\mathbf{D}}_{\lambda\left(  t,\mathbf{x}^{\prime}\right)  }$ commute
for spacelike separated points $\mathbf{x}^{\prime}\mathbf{\neq x}$, so a
measurement at $\mathbf{x}^{\prime}$ does not change the outcome at
$\mathbf{x}$. This is the basis for the description of causal propagation in
QFT. 

\section{Conclusion}

The requirements of QFT described in the Introduction, experimental
observations and the mathematical details of the quantum oscillator model in
Section II are consistent with the following physical picture: In a globally
charge neutral system $\mathbf{D}=\varepsilon_{0}\mathbf{E}+\mathbf{P}%
_{\parallel}$ is transverse and gauge independent and oscillator position,
momentum, mass and external driving force map onto $\mathbf{A}_{\perp}$,
$-\mathbf{D}$, $\varepsilon_{o}$ and $\mathbf{j}_{\perp}$ while $c^{2}%
\mathbf{\nabla\times B=}\Omega^{2}\mathbf{A}_{\perp}$ is the spring-like
restoring force. For any classical field mode second quantization of these
fields ensures that EM excitations are discrete and defines a Fock basis of
$n$-photon states where $n$ is a whole number. Before and after second
quantization these EM fields satisfy MEs and propagate causally in vacuum, a
homogeneous medium and, more generally, any passive optical circuit.
Interaction with polarizable Fermionic matter is described by the local
interaction $-\mathbf{j}_{\perp}\left(  t,\mathbf{x}\right)  \cdot
\mathbf{A}_{\perp}\left(  t,\mathbf{x}\right)  $. Since a one-photon state
collapses to a zero-photon state when the photon is counted, the field
describing it must be interpreted as a probability amplitude. This is verified
in a recent experiment in which a single photon field modeled classically is
delocalize by passage through a biprism. When a detector in placed in each
path, within experimental error, no coincident photon detection events were
observed \cite{BeamSplitter}.

The commutation relations can be used to define a scalar product that can form
a basis for a first quantized theory of single photons. Classical EM fields
are real and covariant, so precise justification of their surprising success
in the interpretation of single-photon experiments \cite{LocalPhotons}
requires first quantized fields that are also real and covariant. Real fields
can be localized in a finite region and "it is well-known to those who know
it" that single-photon interference experiments can be modeled classically
\cite{Barnett}. A justification for this "well know result of Quantum Optics"
is implicit in \cite{Validation} where a real covariant field describing a
single photon is first quantized. The fields can contain can contain a
macroscopic number of photons, making them observable at the time of Faraday
and Maxwell. The CPT theorem explains why the classical Maxwell fields are real.

\end{document}